\documentclass[final,a4paper,12pt,english]{article}

\usepackage{amsmath} 
\usepackage{graphicx}
\usepackage[titletoc]{appendix}
\usepackage{bm}
\usepackage[nomargin,inline,index]{fixme} 
\fxusetheme{color}
\usepackage{color}
\usepackage{ifdraft}                
\usepackage{amssymb}
\usepackage{hyperref} 
\hypersetup{final} 
\usepackage{amsthm}
\usepackage{multirow}
\usepackage{multicol}
\newtheorem{theorem}{Theorem}

\allowdisplaybreaks
\newcommand{\argmin}{\operatornamewithlimits{argmin}}
\newcommand{\ssp}{\mbox{\rm SSP}} \newcommand{\Real}{{\rm
    I}\negthinspace {\rm R}}

\renewcommand{\eqref}[1]{(\ref{#1})}
\newcommand{\secref}[1]{\mbox{\S$\,$\ref{sec:#1}}}

\newcommand{\E}{{\mbox{E}}} 
 \newcommand{\Var}{{\mbox{var}}}
\newcommand{\cum}{{\mbox{cum}}} \newcommand{\Cov}{{\mbox{cov}}}

\newcommand{\thmref}[1]{\mbox{Theorem~\ref{th:#1}}}
\newcommand{\HT}{\mbox{\rm HT}} \newcommand{\sHT}{\mbox{\scriptsize\rm
    HT}} 
\newcommand{\HW}{\mbox{\rm HW}}
\newcommand{\HH}{\mbox{\scriptsize\rm H}}
\newcommand{\sHW}{\mbox{\scriptsize\rm HW}} \newcommand{\MA}{\mbox{\rm
    MA}} \newcommand{\AR}{\mbox{\rm AR}} \newcommand{\ARE}{\mbox{\rm
    ARE}} \newcommand{\ARMA}{\mbox{\rm ARMA}}
\newcommand{\ARFIMA}{\mbox{\rm ARFIMA}}
\newcommand{\pl}{\mbox{\scriptsize\rm PL}}

\title{The Hyv\"arinen scoring rule in Gaussian linear time series
  models}
  \date{}
\author{ Silvia Columbu$^{\rm a}$,  Valentina Mameli$^{\rm b}$, Monica Musio$^{\rm a}$ and A.Philip Dawid$^{\rm c}$  \\
\small{$^{\rm a}$ {\em{Dep. of Mathematics and Computer Science, Univ. of Cagliari, IT}}} \\
\small {$^{\rm b}$ {\em{Dep. of Environmental Sciences, Informatics and Statistics Ca' Foscari Univ.  of Venice, Italy}}} \\
\small {$^{\rm c}$ {\em{Dep. of Pure Mathematics and Mathematical Statistics, Univ. of Cambridge, UK}}}}

\begin{document}

\maketitle
\begin{abstract}
Likelihood-based estimation methods involve the normalising constant
  of the model distributions, expressed as a function of the
  parameter.  However in many problems this function is not easily
  available, and then less efficient but more easily computed
  estimators may be attractive.  In this work we study stationary
  time-series models, and construct and analyse ``score-matching''
  estimators, that do not involve the normalising constant.  We
  consider two scenarios: a single series of increasing length, and an
  increasing number of independent series of fixed length.  In the
  latter case there are two variants, one based on the full data, and
  another based on a sufficient statistic.

  We study the empirical performance of these estimators in three
  special cases, autoregressive (\AR), moving average (MA) and
  fractionally differenced white noise (\ARFIMA) models, and make
  comparisons with full and pairwise likelihood estimators.  The
  results are somewhat model-dependent, with the new estimators doing
  well for $\MA$ and \ARFIMA\ models, but less so for $\AR$ models.

\end{abstract}


\section{Introduction}
\label{intro}
The evaluation of the exact full likelihood may be difficult or even
impossible in situations where complex problems deal with large
correlated datasets.  These problems are likely to occur for instance
in spatial statistics and time series frameworks in which direct
computation of the normalising constant can be a very challenging
task, entailing multidimensional integration of the full joint density
for each value of the parameter.  Many approaches have been proposed
to tackle this issue.  in this paper, we investigate an appealing
approach based on the score matching estimator proposed by Hyv\"arinen
in 2005 \cite{Hyvarinen:2005}.

The score matching estimator can be regarded as a specific case of
estimators derived from proper scoring rules (see Dawid and Musio
(2014) \cite{Dawid:2014}), which are loss functions for measuring the
quality of probabilistic forecasts.  In particular, this estimator
derives from the Hyv\"arinen scoring rule, which is a homogeneous
proper scoring rule (see Ehm and Gneiting (2012) \cite{Ehm:2012} and
Parry {\em et al.} (2012) \cite{Parry:2012}), namely a proper scoring
rule which does not require knowledge of the normalising constant.
Homogeneous scoring rules have been characterised for continuous real
variables (Parry {\em et al.} (2012) \cite{Parry:2012}) and for
discrete variables (Dawid~{\em et al.}  (2012) \cite{Dawid:2012}).  In
a Bayesian framework, Dawid and Musio (2015) \cite{Dawid:Musio:2015}
have shown, for the case of continuous variables, how to handle
Bayesian model selection with improper within-model prior
distributions, by exploiting the use of homogeneous proper scoring
rules.  The discrete counterpart has been empirically studied by Dawid
{\em et al.} (2017) \cite{Dawid:2017}.  In a recent contribution, Shao
{\em et al.} (2018) \cite{Shao:2018} consider the use of the
Hyv\"arinen score for model comparison.  Although the majority of
contributions involving the use of Hyv\"arinen scoring rules focus on
Euclidean spaces, scholars have also investigated extensions to
non-Euclidean spaces: for an early study see Dawid (2007)
\cite{Dawid:2007}.  Recently, Mardia {\em et al.} (2016)
\cite{Mardia:2016} proposed an extension of the Hyv\"arinen scoring
rule to compact oriented Riemannian manifolds, and Takasu {\em et al.}
(2018) \cite{Takasu:2018} constructed a novel class of homogeneous
strictly proper scoring rules for statistical models on spheres.
                 
Given the growing interest in the use of this scoring rule for very
complex statistical models, in this paper we aim to derive an
estimation method based on the Hyv\"arinen scoring rule for estimating
linear Gaussian time series models.

We distinguish two separate cases: a first in which the length of a
single time series increases to infinity, and a second in which the
length of the time series is fixed and the number of series increases
to infinity.

The consistency and asymptotic distribution of the Hyv\"arinen
estimator are derived for the case of a single time series of
increasing length.  In particular, under some mild regularity
conditions we derive consistency of the proposed estimator for linear
Gaussian time series models, while the asymptotic distribution is
proved in the specific case of autoregressive moving average (\ARMA)
causal invertible models.  For time series with fixed length and the
number of time series increasing to infinity the performances of two
estimators based on the Hyv\"arinen scoring rule, namely {\it the
  total Hyv\"arinen estimator} and {\it the matrix Hyv\"arinen
  estimator} are compared through simulation studies with the full
maximum likelihood and the pairwise maximum likelihood estimators.
Three simple time series models have been considered in the design of
simulations: autoregressive (AR), moving average (MA) and fractionally
differenced white noise (ARFIMA).
        
Different behaviours can be detected for the total Hyv\"arinen
estimator among the settings examined.  In particular, it outperforms
the pairwise likelihood estimator in terms of efficiency with respect
to the full maximum likelihood estimator for the MA and ARFIMA
processes.
         
\indent The paper unfolds as follows.  Section~2 introduces basic
notions on scoring rules.  In Section~3 we introduce the Hyv\"arinen
scoring rule for Gaussian linear time series.  Some asymptotic results
for the Hyv\"arinen estimator are given.  In the specific case of $n$
independent series we describe the total Hyv\"arinen estimator and the
matrix Hyv\"arinen estimator.  Section~4 summarises the results of the
simulation studies.  Section~5 presents some concluding remarks.
Technical details are postponed to the Appendices.

\section{Scoring rules}
A {\em scoring rule\/} is a loss function designed to measure the
quality of a proposed probability distribution $Q$, for a random
variable $X$, in light of the outcome $x$ of $X$.  Specifically, if a
forecaster quotes a predictive distribution $Q$ for $X$ and the event
$X=x$ realises, then the forecaster's loss will be $S(x,Q)$.  The
expected value of $S(X,Q)$ when $X$ has distribution $P$ is denoted by
$S(P,Q)$.

The scoring rule $S$ is {\em proper\/} (relative to the class of
distributions $\mathcal{P}$) if
\begin{equation}\label{property}
  S(P,Q)\geq S(P,P),\,\textrm{for all}\, P,\,Q\in\mathcal{P}.
\end{equation}
It is {\em strictly proper\/} if equality obtains only when $Q = P$.

Any proper scoring rule gives rise to a general method for parameter
estimation, based on an unbiased estimating equation: see \secref{est}
below.

\subsection{Examples of proper scoring rules}
Some important proper scoring rules are the log-score,
$S(x,Q)=-\log{q(x)}$ (Good (1952) \cite{Good:1952}), where $q(\cdot)$
is the density function of $Q$, which recovers the full (negative log)
likelihood; and the Brier score (Brier (1950) \cite{Brier:1950}).  A
particularly interesting case, which avoids the need to compute the
normalising constant, produces the {\em score matching\/} estimation
method of Hyv\"arinen (2005) \cite{Hyvarinen:2005}, based on the
following proper scoring rule:
\begin{equation} 
  \label{comphiv} 
  S({\bf x},Q)={\rm \Delta}_{{\bf x}}
  \ln{q({\bf x})} + \frac{1}{2}\left\|\nabla_{{\bf x}} \ln q({\bf x})\right\|^2,
\end{equation}
where ${\bf X}$ ranges over the whole of $\Real^p$ supplied with the
Euclidean norm $\|\cdot\|$, $q(\cdot)$ is assumed twice continuously
differentiable, and $\bf x$ is the realised value of $\bf X$.  In
\eqref{comphiv}, $\nabla_{{\bf x}}$ denotes the gradient operator, and
${\rm \Delta}_{{\bf x}}$ the Laplacian operator, with respect to
$\bf x$.  For $p=1$ we can express
\begin{equation}\label{hyvuni}
  S(x,Q)=\frac{q''(x)}{q(x)} - \frac{1}{2}\left(\frac{q'(x)}{q(x)}\right)^2.
\end{equation}

The scoring rule \eqref{comphiv} is a {\em 2-local homogeneous proper
  scoring rule\/} (see Parry {\em et al.\/} (2012) \cite{Parry:2012}).
It is homogeneous in the density function $q(\cdot)$, {\em i.e.\/} its
value is unaffected by applying a positive scale factor to the density
$q$, and so can be computed even if we only know the density function
up to a scale factor.  Inference performed using any homogeneous
scoring rule does not require knowledge of the normalising constant of
the distribution.

\subsection{Estimation based on proper scoring rules}
\label{sec:est}
Let $(x_1, \ldots, x_{n})$ be independent realisations of a random
variable $X$, having distribution $P_{\theta}$ depending on an unknown
parameter $\theta\in\Theta$, where $\Theta$ is an open subset of
$\Real^m$.  Given a proper scoring rule $S$, let $S(x, \theta)$ denote
$S(x, P_\theta)$.

Inference for the parameter $\theta$ may be performed by minimising
the {\em total empirical score\/},
\begin{equation}
  S(\theta)=\sum_{p=1}^{n} S(x_p,\theta),
\end{equation}
resulting in the {\em minimum score estimator\/},
$\widehat{\theta}_S = \arg\min_{\theta}S(\theta)$.

Under broad regularity conditions on the model (see {\em e.g.\/} Barndorff-Nielsen \& Cox (1994)
\cite{B-N}), $\widehat{\theta}_S$ satistfies the {\em score
  equation\/}:
\begin{equation}
  s(\theta) :=\sum_{p=1}^{n} s(x_p, \theta)=0,	
\end{equation} 
where $s(x, \theta) := \nabla_{\theta} S(x,\theta)$, the gradient
vector of $S(x,\theta)$ with respect to $\theta$.  The score equation
is an unbiased estimating equation (Dawid \& Lauritzen (2005)
\cite{Dawid:2005}).  When $S$ is the log-score, the minimum score
estimator becomes the maximum likelihood estimator.

From the general theory of unbiased estimating functions, under broad
regularity conditions on the model the minimum score estimate
$\widehat{\theta}_S$ is asymptotically consistent and normally
distributed:
$\widehat{\theta}_S \sim N(\theta, \left\{n G(\theta)\right\}^{-1}),$
where $G(\theta)$ denotes the {\em Godambe information matrix\/}
$G(\theta):=K(\theta)J(\theta)^{-1}K(\theta),$ where
$J(\theta)=\E\left\{ s(X,\theta)s(X,\theta)^T\right\}$ is the {\em
  variability matrix\/}, and
$K(\theta)=\E\left\{\nabla s(X,\theta)^T\right\}$ is the {\em
  sensitivity matrix\/}.  In contrast to the case for the full
likelihood, $J$ and $K$ are different in general: see
Dawid \& Musio (2014) \cite{Dawid:2014}, Dawid {\em et al} (2016)\cite{Dawid:2016}.  We point out that estimation of
the matrix $J(\theta)$, and (to a somewhat lesser extent) of the
matrix $K(\theta)$, is not an easy task: see Varin (2008)
\cite{Varin:2008}, Varin {\em et al.} (2011) \cite{Varin:2011} and
Cattelan and Sartori (2016) \cite{Cattelan}.

\section{Gaussian linear time series models}\label{models}
In this section we introduce some results based on the use of the
Hyv\"arinen scoring rule in the setting of Gaussian linear time series
models.

Consider the Gaussian linear time series model
\begin{equation} \label{lts} y_t=\mu+\sum_{j=0}^{\infty}\psi_jz_{t-j},
  \quad t=1,2,\ldots,
\end{equation}
parametrised by $\mu$, $\sigma^2$ and $\lambda\in\Real^{m-2}$, where
for $j\geq 0$, $\psi_j=\psi_j(\lambda)$ satisfies $\psi_0=1$ and
$\sum_{t=0}^{\infty}\psi_t^2<\infty$.  The $(z_{t})$ are iid Gaussian
variables with mean $0$ and variance $\sigma^2$.  Let
$\theta = (\mu, \sigma^2,\lambda)$ be the vector of model parameters.
The autocovariance function is
$\E\{(y_{t+j}-\mu)(y_t-\mu)\}=\sigma^2\sum_{t=0}^{\infty}\psi_t\psi_{t+j}=\sigma^2\gamma_{\lambda}(j)$,
say.
Using basic differentiation rules, it is easy to find the Hyv\"arinen
score based on the single time series $Y_T=(y_1,y_2,\ldots,y_T)$:
\begin{eqnarray}\label{hivunima}
  H(\theta,Y_T)&=&-\frac{1}{\sigma^2}\sum_{i=1}^T{\rm\Gamma}^{ii}+\frac{1}{2}\sum_{i=1}^T
                   \left\{ \sum_{t=1}^T\frac{1}{\sigma^2}{\rm\Gamma}^{it}(y_t-\mu)\right\} ^2,
\end{eqnarray}
where the matrix ${\rm\Gamma}$ has $(i,j)$ entry
${\rm\Gamma}_{ij}=\gamma_{\lambda}(|i-j|)$ and ${\rm\Gamma}^{ij}$ is
the $(i,j)$ entry of ${\rm\Gamma}^{-1}$.
We will denote denote the Hyv\"arinen estimator based on a single series
by $\widehat{\theta}_{\HH}$.


\subsection{Asymptotic results for a single time series}

\label{sec:4}
In this section we analyse the asymptotic statistical properties of
the Hyv\"arinen scoring rule estimator, based on \eqref{hivunima}, for
a single time series.

The following theorem shows the consistency of the estimator
$\widehat{\theta}_{\HH}$ in the Gaussian linear time series setting.  The
proof of the Theorem is deferred to Appendix and follows
arguments similar to those used by Davis and Yau (2011) \cite{Davis}
to prove the consistency of the pairwise likelihood estimator.
\begin{theorem}\label{th:1}
  Suppose $(y_t)$ is the linear process in \eqref{lts} with $\mu=0$
  and parameter $\theta_0=(\sigma^2_0, \lambda_0)$.  Let
  \[\widehat{\theta}_{\HH}=\argmin_{\theta} H(\theta, Y_T)\]
  be the minimum score estimator, where $\theta=(\sigma^2,\lambda)$ and $\lambda \in \Lambda$, where $\Lambda$ is a compact set.  If the identifiability condition
  \begin{equation} \label{identif} \sigma_1^2 \gamma_{\lambda_1}(j)=
    \sigma_2^2 \gamma_{\lambda_2}(j) \text{ for } j=0,1, \ldots, k
    \text{ iff } (\sigma^2_1, \lambda_1)=(\sigma^2_2, \lambda_2)
  \end{equation}
  is satisfied, then $\widehat{\theta}_{\HH} \xrightarrow{a.s.} \theta_0$
  as $T\rightarrow \infty$.
\end{theorem}

Once consistency has been proved, we focus on the asymptotic
distribution of $\widehat{\theta}_{\HH}$.  Its analytic form involves the
elements $\Gamma^{ij}$ of the inverse of the autocovariance matrix.
In order to guarantee its absolute summability, we restrict our
attention to the case of ARMA causal invertible processes.

Defining $b_{ij}={\Gamma^{ij}}/{\sigma^2}$, the gradient and the
Hessian with respect to $\widehat{\theta}_{\HH}$ are given, respectively,
by the following two expressions:
\begin{eqnarray}
  \label{eq:j}
  J(\theta)=\frac{\partial}{\partial \theta} H(\theta, Y_T) &=&
                                                                -\sum_{i=1}^T \frac{\partial b_{ii}}{\partial \theta}+ \frac{2}{2}
                                                                \sum_{i,j,t=1}^T \frac{\partial b_{ij}}{\partial \theta}b_{it} y_j
                                                                y_t\\
\nonumber  K(\theta)= \frac{\partial^2}{\partial \theta^2} H(\theta, Y_T) &=&
                                                                     -\sum_{i=1}^T \frac{\partial^2 b_{ii}}{\partial \theta^2}+
                                                                     \sum_{i,j,t=1}^T \frac{\partial b_{ij}}{\partial \theta}
                                                                     \frac{\partial b_{it}}{\partial \theta}y_j y_t\\
  \label{eq:k}                                                            &&\mbox{} + \sum_{i,j,t=1}^T
                                                                             \frac{\partial^2 b_{ij}}{\partial \theta^2} b_{it} y_j y_t
\end{eqnarray}
where ${\partial}/{\partial \theta}$ denotes differentiation with
respect to the components of the vector $\theta$.

\begin{theorem}\label{th:2}
  Suppose that $(y_t)$ is an $\ARMA(p,q)$ causal and invertible
  process.  Furthermore, assume that the Hessian matrix $K(\theta)$ is
  invertible in a neighbourhood of $\theta_0$.  If the identifiability
  condition \eqref{identif} holds, then
  \[\sqrt{T}(\widehat{\theta}_T-\theta_0)\xrightarrow{\mathcal{D}}
    N_{m-1} \left(0, K(\theta_0)^{-1} V K^T(\theta_0)^{-1}\right),\]
  where
  $V=\sum_{r=-\infty}^{\infty} \sum_{k=-\infty}^{\infty}
  V(r,k,\theta_0)$ with
  $$V(r,k,\theta_0)= \left(\frac{\partial }{\partial
      \theta_0}\frac{\gamma^{-1}(0)}{\sigma^2_0} \right)^2+
  \frac{\partial }{\partial \theta_0}
  \frac{\gamma^{-1}(k)}{\sigma_0^2} \gamma^{-1}(0) \frac{\partial
    \gamma^{-1}(k+r)}{\partial \theta_0} \gamma(r).$$
\end{theorem}
\thmref{2} shows that the Hyv\"arinen scoring rule estimator
$\widehat{\theta}_{\HH}$, in the case that $(y_t)$ is an ARMA causal
invertible process, is asymptotically normally distributed with rate
of decay $\sqrt{T}$.  As is well known, the autocovariance function of
an ARMA process decays exponentially, which means that an ARMA process
is a short memory process, and its autocovariance function is
absolutely summable Brockwell \& Davis (1991) \cite{Brockwell:1991}.  This property, together
with the duality of ARMA models under causality and invertibility,
allows us to prove asymptotic normality. For the complete proof refer to the appendix.


\subsection{Estimation approaches for $n$ independent time
  series}\label{mhyv}
In the remainder of this section we discuss the case of $n$
independent series of length $T$.  We assume that $T$ is fixed while
$n$ increases to infinity.  We also specialise to the case that the
common mean $\mu$ and innovation variance $\sigma^2=\sigma_0^2$ are
known; without loss of generality we take $\mu=0$.  

Consider now $n$ independent and identically distributed processes
$Y_{1}, \ldots ,Y_{n}$, where $Y_p = (y_{p1},\ldots,y_{pT})$, each
having the $T$-variate normal distribution with mean-vector $0$ and
variance covariance matrix $\sigma^2{\rm\Gamma}$, with unknown
parameter $\lambda$.  Let the $(n\times T)$ random matrix $Y$ have the
vector $Y_{p}$ as its $p$th row.  We define the {\it total
  Hyv\"{a}rinen score} (\HT) as the sum of $n$ single Hyv\"{a}rinen
scores in \eqref{hivunima}:

\begin{equation} \label{HT} \HT(\lambda)= \sum_{p=1}^n H_p
  (\lambda,Y_{p}),
\end{equation}
where
\begin{eqnarray}
  H_p(\lambda,Y_{p})&=&-\frac{1}{\sigma^2}\sum_{i=1}^T{\rm\Gamma}^{ii}+\frac{1}{2}\sum_{i=1}^T
                        \left\{ \sum_{t=1}^T\frac{1}{\sigma^2}{\rm\Gamma}^{it}y_{pt}\right\} ^2.
\end{eqnarray}
The estimate of $\lambda$ minimising the total Hyv\"{a}rinen score
will be denoted by $\widehat{\lambda}_{\sHT}$.

An alternative approach is to consider as basic observable the
sum-of-squares-and-products matrix $\ssp=Y^TY$, which is a sufficient
statistic for the multivariate normal model, having the Wishart
distribution with $n$ degrees of freedom and scale matrix
$\sigma^2{\rm\Gamma}$.  Then inference for the parameter $\lambda$ can
be performed by resorting to the Hyv\"{a}rinen score based directly on
the Wishart model.  We will call this scoring rule the \textit{matrix
  Hyv\"{a}rinen score}.

Assuming $n\geq T$, so that the joint distribution of the upper
triangle $\left(s_{ij}: 1\leq i\leq j\leq T\right)$ of the
sum-of-squares-and-products random matrix $\ssp$ (which has a Wishart
distribution with parameters $n $ and $\sigma^2{\rm\Gamma}$) has a
density, and taking into consideration all of the properties of the
derivatives of traces and determinants, it can be shown that the
Hyv\"arinen score based on this joint density is
\begin{equation}\label{HS}
  \HW(\ssp,{\rm\Gamma})=-\frac{(n-T-1)}{2}\sum_{i=1}^{T}(s^{ii})^2+\frac{1}{2}\sum_{i,j=1}^{T}
  \left\{\frac{(n-T-1)}{2}s^{ij}-\frac{1}{2\sigma^2}{\rm\Gamma}^{ij}\right\}^2,
\end{equation}
\noindent where $s^{ij}$ and ${\rm\Gamma}^{ij}$ are the elements of
the inverse matrices $\ssp^{-1}$ and of ${\rm\Gamma}^{-1}$,
respectively.  The matrix Hyv\"arinen estimator for $\lambda$,
minimising $\HW(\ssp,{\rm\Gamma})$ with respect to $\lambda$, will be
denoted by $\widehat\lambda_{\sHW}$.

The derivative of $\HW(\ssp,{\rm\Gamma})$ with respect to $\lambda$ is
\begin{equation}\label{der}
  \HW_{\lambda}(\ssp,{\rm\Gamma})=-\frac{1}{2\sigma^2}\sum_{i,j=1}^{T}\left\{\frac{(n-T-1)}{2}s^{ij}-\frac{1}{2\sigma^2}
    {\rm\Gamma}^{ij}\right\}\frac{\partial{\rm\Gamma}^{ij}}{\partial\lambda},
\end{equation}
\noindent and $\E\left\{\HW_{\lambda}(\ssp,{\rm\Gamma})\right\}=0$
since $\E\left( s^{ij}\right)={{\rm\Gamma}^{ij}}/(\sigma^2(n-T-1)) $
(see Kollo \& von Rosen, p.~257).  Moreover,
$K(\lambda)=\E\left\{
  \HW_{\lambda\lambda}(\ssp,{\rm\Gamma})\right\}=\sum_{i,j=1}^{T}
\left({\partial{\rm\Gamma}^{ij}}/{\partial\lambda}\right)^2/4\sigma^4$.
The derivation of the function $J(\lambda)$, which after taking
account of the square of \eqref{der} reduces to
\begin{equation}
  J(\lambda)=\frac{(n-T-1)^2}{16\sigma^4}\sum_{i,j,k,l=1}^{T}\left( \frac{\partial{\rm\Gamma}^{ij}}
    {\partial\lambda}\right) ^2\mbox{cov}\left( s^{ij},s^{kl}\right),
\end{equation} 
involves calculations requiring the covariance matrix of the random
matrix $\ssp^{-1}$, which has an Inverse Wishart distribution with
scale matrix $\frac{1}{\sigma^2}{\rm\Gamma}^{-1}$: see Von Rosen (1998) \cite{Rosen}
for details on the components of the covariance matrix.

In general, the Godambe information needed to estimate the standard
error of $\widehat{\lambda}_{\sHW}$ is not easy to derive analytically
due to the form of the matrix ${\rm\Gamma}$.  It should be pointed out
that this approach can not be used if we have a single time series of
length $T$ with $T$ increasing to $\infty$, since for non-singularity
of the Wishart distribution we need to assume $n\geq T$.


\section{Numerical assessment}
\label{sec:5}
In this section we report simulation studies designed to assess and
compare the behaviours of the estimators obtained by using the total
and the matrix Hyv\"arinen estimators.  We refer to the case described
in paragraph \ref{mhyv} in which $T$ is fixed and $n$ increases to
$\infty$.  For comparison, we will consider also the full and pairwise
maximum likelihood estimators (Davis \& Yau (2011) \cite{Davis} ).  We
discuss three examples: the first order autoregressive AR(1), the
first order moving average \MA(1) and the fractionally differenced
white noise $\ARFIMA(0,d,0)$.  Various parameter settings are
considered in all simulation studies.  All calculations have been done
in the statistical computing environment {\tt R} (\cite{R}).  The
summary statistics shown are: average estimates of the parameters,
asymptotic standard deviations ($sd$) and the asymptotic relative
efficiency (ARE) with respect to the maximum likelihood estimator.

\subsection{First order autoregressive models}\label{ar1}
The stationary univariate autoregressive process of order $1$, denoted
by $\AR(1)$, is defined by
\begin{eqnarray*}
  y_{1} &=&  \mu +\frac{1}{\sqrt{1-\phi^2}}\, z_{1}\\
  y_{t}  &=& \mu+\phi(y_{t-1} -\mu) +z_{t}, \quad (t=2,\dots,T), 
\end{eqnarray*}
where $(z_{t})$ is a Gaussian white noise process with mean $0$ and
variance $\sigma^2$.  Here $\phi$, with $|\phi|<1$, is the {\em
  autoregressive parameter\/}.  Then $y_{1},\ldots,y_{T}$ are jointly
normal with mean vector $\mu1_T$ (where $1_T$ is the $T$-dimensional
unit vector), and covariance matrix $\sigma^2{\rm\Gamma}$, with
$\Gamma$ having components
${\rm\Gamma}_{lm}={\phi^{|l-m|}}/{(1-\phi^2)}$ ($l, m = 1,\ldots, T$).

For comparison purposes we consider also the numerical performance of
a class of pairwise likelihood estimators.  Since, in the time series
considered, dependence decreases in time, as in Davis \& Yau (2011)
\cite{Davis} we shall restrict to the {\em first order consecutive
  pairwise likelihood\/}, rather than the complete pairwise
likelihood, so that adjacent observations are more closely related
than the others.  This choice is motivated also by the loss in
efficiency incurred in using the $k$-th order consecutive pairwise
likelihood as $k$ increases (see Davis and Yau (2011) \cite{Davis};
Joe and Lee (2009) \cite{Joe}).  Note that, when it is known that
$\mu=0$ but $\sigma^2$ is unknown, the pairwise likelihood estimator
of $\phi$ is
$ \widehat{\phi}_{\pl} =
2\sum_{t=2}^{T}y_ty_{t-1}/\sum_{t=2}^{T}(y_t^2+y_{t-1}^2)$, which is
also the Yule-Walker estimator (Davis \& Yau (2011) \cite{Davis}).

\begin{figure}
  \includegraphics[scale=0.5,keepaspectratio]{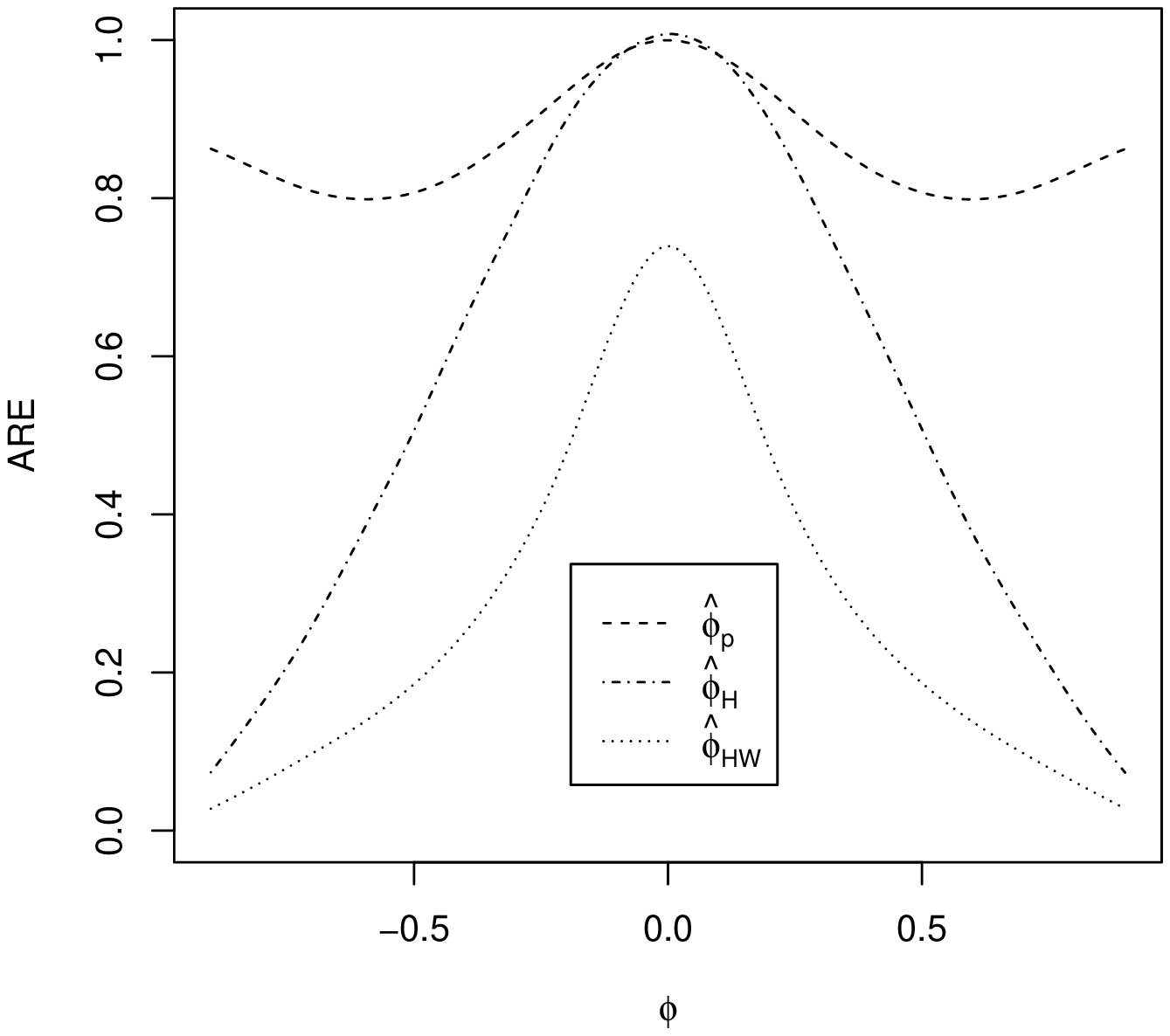}
  \includegraphics[scale=0.5,keepaspectratio]{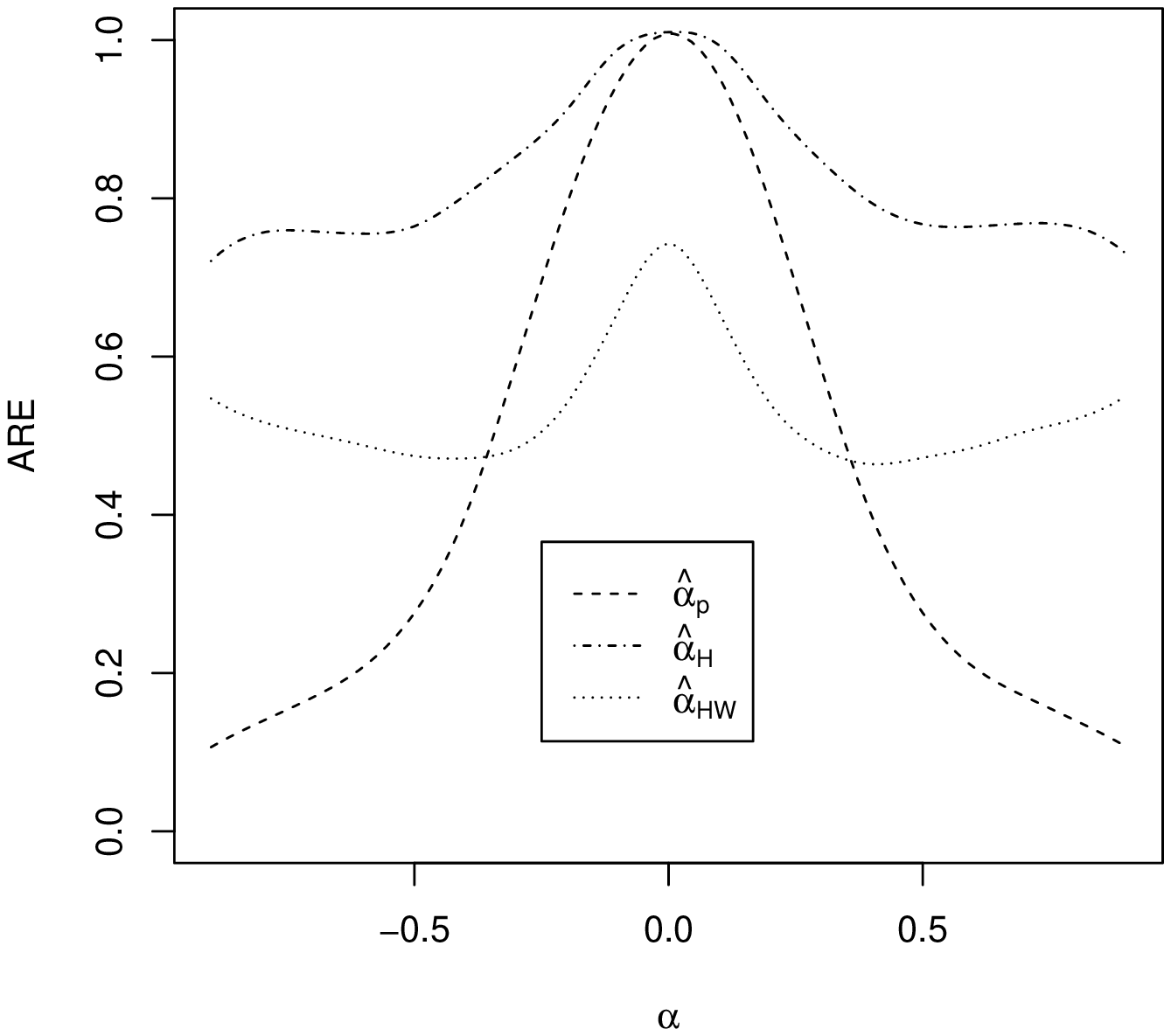}
  \includegraphics[scale=0.5,keepaspectratio]{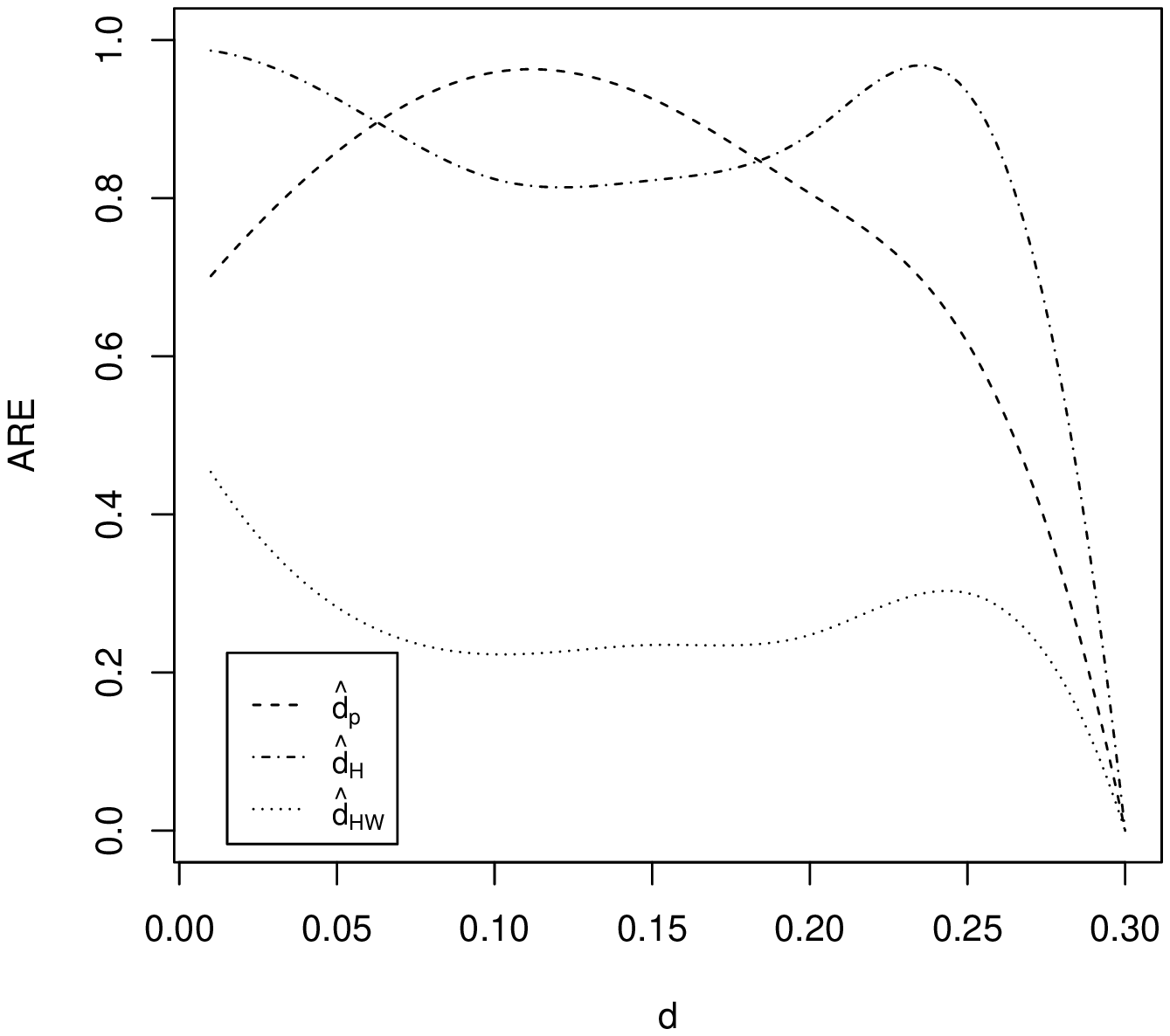}
  \caption{Asymptotic relative efficiency of estimators for the
    $\AR(1)$ model (left top panel), for the $\MA(1)$ model (right top panel)
    and for the $\ARFIMA(0,d,0)$ model (left bottom panel).}
  \label{fig1}
\end{figure}

\paragraph{Simulation~1}
The values of the model parameters are $\mu=0$ and $\sigma=1$, with
the autoregressive parameter $\phi\in\{-0.9,-0.8,\ldots,0.8,0.9\}$.
In the simulation study, $1000$ replicates are generated of $n = 200$
processes of length $T=50$.  Results are summarised in
Table~\ref{tab1}.  The numerical results in Table~\ref{tab1} and in
the left-hand panel of Figure~\ref{fig1} suggest that
$\widehat{\phi}_{\sHT}$ and $\widehat{\phi}_{\sHW}$ do not have high
efficiency as $\phi$ approaches $1$: in particular, the asymptotic
efficiency of $\widehat{\phi}_{\sHW}$ tends to $0$ for large values of
$|\phi|$.  In contrast, under the same model setting, there is only a
modest loss of efficiency for the pairwise likelihood-based estimator
$\widehat{\phi}_{\pl}$.
\subsection{First order moving average models}
The univariate moving average process of order $1$, denoted by
$\MA(1)$, is defined by
\begin{equation}
  y_{t}= \mu + \alpha z_{t-1} +z_{t},\quad \quad (t=1,\dots,T),
\end{equation}
where $|\alpha|<1$ and $z_0,\ldots,z_T$ are independent Gaussian
random variables with $0$ mean and variance $\sigma^2$.  Then the
random variables $y_{1},\ldots,y_{T}$ are jointly normal, each having
mean $\mu$ and variance $\sigma^2(1+\alpha^2)$.  The variables $y_t$
and $y_{t+k}$ are independent for $|k| > 1$, while $y_t$ and $y_{t+1}$
have covariance $\sigma^2\alpha$ ($t= 1,\ldots, T-1$).  Hence, the
covariance matrix $\sigma^2{\rm\Gamma}$ of $Y=(y_1,y_2,\ldots,y_T)$
has components $\sigma^2{\rm\Gamma}_{ss}=\sigma^2(1+\alpha^2)$,
$\sigma^2{\rm\Gamma}_{st}=\sigma^2\alpha$ if $|s-t|=1$,
$\sigma^2{\rm\Gamma}_{st}=0$ otherwise.

As before we consider the first order consecutive pairwise likelihood
since the use of a higher order consecutive pairwise likelihood is
unrealistic due to the independence of $y_t$ and $y_{t+k}$ for
$k\geq 2$.  For \mbox{$t=1,\ldots,T-1$}, the pair $(y_t,y_{t+1})$ has
a bivariate Gaussian distribution, in which the two components both
have mean $\mu$ and variance $\sigma^2(1+\alpha^2)$, and have
covariance $\sigma^2\alpha$.

\paragraph{Simulation~2}
The values of the model parameters are $\mu=0$ and $\sigma=1$, with
the moving average parameter $\alpha\in\{-0.9,-0.8,\ldots,0.8,0.9\}$.
In the simulation study, $1000$ replicates are generated of $n = 200$
processes of length $T=50$.  Results are summarised in
Table~\ref{tab2}.  The simulation shows that the total Hyv\"arinen
estimator $\widehat{\alpha}_{\sHT}$ achieves the same efficiency as the
MLE in the $\MA(1)$ model for values of the moving average parameter
near $0$; see Table~\ref{tab2} and the right-hand panel of
Figure~\ref{fig1}.  However, the loss in efficiency of the total
Hyv\"arinen estimator $\widehat{\alpha}_{\sHT}$ is modest even when the
absolute value of the moving average parameter reaches $1$.  In
contrast, the pairwise likelihood estimator $\widehat{\alpha}_{\pl}$
shows very poor performances in terms of asymptotic relative
efficiency: the ARE ranges from $1$ to $0.1$ as $|\alpha|$ increases.

\subsection{Fractionally differenced white noise}
The fractionally differenced white noise, $\ARFIMA(0,d,0)$, model is
defined by
\begin{equation*}
  (1-\Pi)^dy_t = z_t,\, \textrm{with}\, t=1,\dots,T,
\end{equation*}
where $\Pi$ is the lag operator and $d \in (0, 0.5)$, and
$z_1,\ldots,z_T$ are independent Gaussian random variables with $0$
mean and variance $\sigma^2$.  Then the random variables
$y_{1},\ldots,y_{T}$ are jointly normal, with covariance matrix
$\sigma^2{\rm\Gamma}$ whose components (see Hosking (1981) \cite{Hosking}) are
\begin{equation}
  \label{eq:hosk}
\sigma^2{\rm\Gamma}_{ij}=\frac{(-1)^{|k|}\sigma^2{\rm\Gamma}(1-2d)}
{{\rm\Gamma}(|k|-d+1){\rm\Gamma}(-|k|-d+1)}\qquad(k=i-j)  
\end{equation}
(where in the right-hand side of \eqref{eq:hosk}, $\Gamma$ denotes the
gamma function.)

As before we consider the first order consecutive pairwise likelihood
since no great improvement can be detected by using a higher order
consecutive pairwise likelihood: see the results of Davis and Yau
(2011) \cite{Davis}.  For \mbox{$t=1,\ldots,T-1$}, the pair
$(y_t,y_{t+1})$ has a bivariate Gaussian distribution, in which the
two components both have mean $\mu$ and variance
$\sigma^2{{\rm\Gamma}(1-2d)}/{{\rm\Gamma}(1-d)^2}$, and have
covariance $-\sigma^2{{\rm\Gamma}(1-2d)}/{{\rm\Gamma}(2-d){\rm\Gamma}(-d)}$.

\paragraph{Simulation~3}
The values of the model parameters are $\mu=0$ and $\sigma=1$, with
the fractional parameter $d\in\{0.01,0.05,0.1,0.15,0.2,0.25,0.3\}$.
In the simulation study, $1000$ replicates are generated of $n = 100$
processes of length $T=50$.  Results are sumarised in
Table~\ref{tab3}.  Simulation~3 shows that the total Hyv\"arinen
estimator $\widehat{d}_{\sHT}$ achieves the same efficiency as the MLE
in the $\ARFIMA(0,d,0)$ model near $0$ and near $0.3$; see
Table~\ref{tab3} and the right-hand panel of Figure~\ref{fig1}.  The
loss in efficiency of the total Hyv\"arinen estimator
$\widehat{d}_{\sHT}$ is very slight when $d\in (0,0.3)$.  The
efficiency of $\widehat{d}_{\sHW}$ is poor with ARE values ranging
from $0$ to $0.45$.  For all the estimators considered the ARE is 0
when $d\in (0.3,0.5)$.  The pairwise estimator $\widehat{d}_{\pl}$
performs better than $\widehat{d}_{\sHW}$, however the values of ARE
range from 0.6 to 0.96, reaching a maximum when $d=0.1$, with a major
loss of efficiency with respect to the total Hyv\"arinen estimator.
\subsection{Discussion}
It should be noted that for the $\MA(1)$ and the $\ARFIMA(0,d,0)$
models no analytic expressions for the derivatives of \eqref{hivunima}
are available.  The standard deviations of $\widehat{\phi}_{\sHT}$,
$\widehat{\alpha}_{\sHT}$ and $\widehat{d}_{\sHT}$ are empirical
estimates of the square root of the Godambe information function,
which is obtained by compounding the empirical estimates of $J$ and
$K$.  The standard deviations of the pairwise maximum likelihood
estimator and the maximum likelihood estimator are obtained using the
analytic expressions (see Pace et al. (2011) \cite{Pace}) for the $\AR(1)$ model and the
empirical counterparts for the $\MA(1)$ model.  Numerical evaluation
of scoring rule derivatives has been carried out using the {\tt R}
package {\tt numDeriv}; see Gilbert \& Varadhan (2012) \cite{num}.

Results from simulations reveal that the estimators considered produce
estimates very close to the true values of the parameters.  However,
results not shown here suggest that when the length $T$ of the series
is small the pairwise likelihood estimator performs worse in terms of
bias than the other estimators in all the models considered.

The left, the middle and right-hand panels of Figure~\ref{fig1} depict
the asymptotic relative efficiency as a function of $\phi$ for the
$\AR(1)$ model, as a function of $\alpha$ for the $\MA(1)$ model, and
as a function of $d$ for the $\ARFIMA(0,d,0)$ model, respectively.

All the results of the simulation studies are in agreement with the
findings of Davis \& Yau (2011) \cite{Davis} who focus on pairwise likelihood-based
methods for linear time series.

\section{Conclusions}

In this paper we have considered the use of Hyv\"arinen scoring rules
in linear time series estimation under different conditions.  We have
established the consistency of the Hyv\"arinen scoring rule estimator
for a single times series under some general conditions and its
asymptotic normality in an ARMA time series context.
 
We have investigated, for $n$ independent time series, the
performances of two estimators based on the Hyv\"arinen scoring rule,
which can be regarded as a surrogate for a complex full likelihood.
The properties of the estimators found using this scoring rule are
compared with the full and pairwise maximum likelihood estimators.
Three simple models are discussed: the first a stationary first order
autoregressive model, the second a first order moving average model
and the third a fractionally differenced white noise.  In the first
case the total Hyv\"arinen method leads to poor estimators; in
contrast, in the second and third this method produces good
estimators.  The opposite behaviour is observed for the pairwise
estimators.  For the moving average process and the fractionally
differenced white noise, there can be a large gain in efficiency, as
compared to the pairwise likelihood method, by using the total or the
matrix Hyv\"arinen scoring rule estimators.  For the autoregressive
model, in contrast, the total Hyv\"arinen score methods suffer a loss
of efficiency as $|\phi|$ approaches $1$.

In all examples, results not reported here show that there is a great
improvement in the performances of the matrix Hyv\"arinen estimator
based on the Wishart model as the ratio $T/n$ becomes negligible.  It
is clear that the loss of efficiency incurred in using the Hyv\"arinen
scoring rules or pairwise likelihood can be substantial, but this
depends on the underlying model, and no overall general principle has
emerged that might offer guidance for cases not yet considered.  The
matrix Hyv\"arinen estimator has the apparent advantage over the other
estimators (apart from full maximum likelihood) of being based on the
sufficient statistic of the model; nevertheless the total Hyv\"arinen
estimator shows good performance in terms of efficiency.

We conclude that minimising the total Hyv\"arinen score may offer a
viable and useful approach to estimation in models where computation
of the normalising constant in the likelihood function is not
feasible, and pairwise likelihood methods lead to poor estimators.

\section*{Acknowledgements}
Monica Musio was supported by the project GESTA of the Fondazione di
Sardegna and Regione Autonoma di Sardegna.
\section*{Appendix: Proof of Theorem \ref{th:1}}
Let $\theta=(\sigma^2, \lambda)$ and let $\E_{\theta}$ denote the
expectation with respect to the probability distribution for $(y_t)$
defined in Equation~\eqref{lts}.  Let
$\theta_0 = (\sigma^2_0, \lambda_0)$ denote the true parameter value.
From the ergodicity of $(y_t)$, it follows that $H(\theta, Y_T)$ is
ergodic and stationary and therefore
  \begin{equation}\label{Hlim}
    \frac{1}{T} H(\theta, Y_T) \xrightarrow{a.s.} H(\theta_0,\theta) := \E_{\theta_0} H(\theta, y_1).
  \end{equation}
  Since the Hyv\"arinen score is strictly proper we have
  \begin{equation} 
    \label{inequality} H(\theta_0,\theta)\geq  H(\theta_0,\theta_0)
  \end{equation}
  with equality if and only if $\theta=\theta_0$, by the
  identifiability condition \eqref{identif}.
The approach used to derive the consistency of the total Hyv\"arinen
  estimator now follows the same general argument used to derive the
  consistency of the pairwise likelihood estimator in Davis \& Yau (2011) \cite{Davis}.
  
  In particular, the compactness of $\Lambda$ and the inequality
  \eqref{inequality} are used as devices for proving the claim.

\section*{Appendix: Proof of Theorem \ref{th:2}}
Define the sample gradient and Hessian as
\[
  J_T(\theta) := -\frac{1}{T}\sum_{i=1}^T \frac{\partial
    b_{ii}}{\partial \theta}+ \frac{2}{2T} \sum_{i,j,t=1}^T
  \frac{\partial b_{ij}}{\partial \theta}b_{it} y_j y_t
\]
and
\[
  K_T(\theta) := -\frac{1}{T}\sum_{i=1}^T \frac{\partial^2
    b_{ii}}{\partial \theta^2}+ \frac{1}{T}\sum_{i,j,t=1}^T
  \frac{\partial b_{ij}}{\partial \theta} \frac{\partial
    b_{it}}{\partial \theta}y_j y_t+ \frac{1}{T}\sum_{i,j,t=1}^T
  \frac{\partial^2 b_{ij}}{\partial \theta^2} b_{it} y_j y_t.
\]

Using a Taylor expansion of $J_T(\theta)$ around $\theta_0$ and the
consistency of Hyv\"arinen scoring rule estimator, it can be proved
that, for some $\theta_T^+$ between $\theta_0$ and
$\widehat{\theta}_T$,
\begin{equation}\label{grad}
  J_T(\theta_0)= K_T(\theta_T^+) \sqrt{T}(\theta_0-\widehat{\theta}_T).
\end{equation}

The asymptotic distribution of $\widehat{\theta}_T$ can be derived by
exploiting the asymptotic properties of $K_T(\theta_T^+)$ and
$J_T(\theta_0)$, together with the ergodic theorem and the fact that
$\theta^+_T \xrightarrow{a.s.} \theta_0$.

Writing
\begin{eqnarray*}
  \frac{\partial}{\partial\theta_0} &=&\left.\frac{\partial} {\partial\theta}\right|_{\theta=\theta_0}\\
  \Gamma &=& \Gamma(\lambda_0)
\end{eqnarray*}
it can be shown that
\[
  K_T(\theta_T^+) \xrightarrow{a.s.} \sum_{i,j,t=1}^T \frac{\partial
    b_{ij}}{\partial \theta_0} \frac{\partial b_{it}}{\partial
    \theta_0} \sigma^2_0 \Gamma_{tj}= K(\theta_0).
\]
In order to calculate the asymptotic distribution of
$\widehat{\theta}_T$ we need to calculate the expectation and the
variance of $J_T(\theta_0)$.  The calculation of the expectation of
$J_T(\theta_0)$ follows easily from the unbiasdness of the scoring
rule estimating equation (Dawid \& Lauritzen  (2005) \cite{Dawid:2005}).  However,
calculation of the variance of $J_T(\theta_0)$ is challenging due to
the presence of the non deterministic term
\[
  B_{i}= \sum_{j,t=1}^T \frac{\partial b_{ij}}{\partial
    \theta_0}b_{it} y_j y_t.
\]
It relies on the following calculation:
\begin{eqnarray}\label{eq_}
  \Var(J_T(\theta_0))&=&\frac{1}{T} \sum_{i=1}^T \Var( B_{i})=\nonumber\\
                     &=&\frac{1}{T} \sum_{i,j,t,\ell,h=1}^T   \frac{\partial b_{ij}}{\partial \theta_0}b_{it}  
                         \frac{\partial b_{i\ell}}{\partial \theta_0}b_{ih} \Cov(y_j y_t, y_\ell y_h) \nonumber \\
                     &=&\sum_{i,j,t,\ell,h=1}^T   \frac{\partial b_{ij}}{\partial \theta_0}\frac{\Gamma^{it}}{\sigma_0^2} 
                         \frac{\partial b_{i\ell}}{\partial \theta_0}\frac{\Gamma^{ih}}{\sigma_0^2} \{\Cov(y_j , y_\ell) \Cov(y_t , y_h)  \nonumber \\
                     &&\mbox{}+ \Cov(y_j , y_h) \Cov(y_t , y_\ell) 
                        +  \cum_4(y_j, y_t, y_\ell, y_h)\}\nonumber\\
                     &=&\sum_{i,j,t,\ell,h=1}^TA_{j\ell th}+C_{jht\ell}+  D_{it\ell h},
\end{eqnarray}
where 
\begin{eqnarray*} 
A_{j\ell th}&=&\frac{\partial b_{ij}}{\partial \theta_0}\frac{\Gamma^{it}}{\sigma_0^2} 
\frac{\partial b_{i\ell}}{\partial \theta_0}\frac{\Gamma^{ih}}{\sigma_0^2} \Cov(y_j , y_\ell) \Cov(y_t , y_h)\\
C_{jht\ell}&=&\frac{\partial b_{ij}}{\partial
  \theta_0}\frac{\Gamma^{it}}{\sigma_0^2} \frac{\partial
  b_{i\ell}}{\partial \theta_0}\frac{\Gamma^{ih}}{\sigma_0^2} \Cov(y_j
, y_h) \Cov(y_t , y_\ell)\\
D_{it\ell h}&=&\frac{\partial b_{ij}}{\partial
  \theta_0}\frac{\Gamma^{it}}{\sigma_0^2} \frac{\partial
  b_{i\ell}}{\partial \theta_0}\frac{\Gamma^{ih}}{\sigma_0^2}
\cum_4(y_j, y_t, y_\ell, y_h).
\end{eqnarray*}

The first term in \eqref{eq_} simplifies as
\begin{align}\label{first}
  \sum_{i,j,t,\ell,h=1}^T   A_{j\ell th} = \sum_{i,j,t,\ell,h=1}^T   \frac{\partial b_{ij}}
  {\partial \theta_0}\Gamma^{ii} \frac{\partial b_{i\ell}}{\partial \theta_0} \Gamma_{j \ell}.
\end{align}
The second term simplifies as
\begin{align}\label{second}
  \sum_{i,j,t,\ell,h=1}^T  C_{jht\ell}=T\left( \frac{\partial} 
  {\partial \theta_0} \frac{\gamma^{-1}(0)}{\sigma_0^2}\right)^2.
\end{align}
The third term in \eqref{eq_}, which involves the fourth cumulant,
vanishes as for Gaussian linear processes all the cumulant functions
$\cum_k$ for $k>3$ are identically null Brockwell \& Davis (1991) \cite{Brockwell:1991}.  Hence
convergence of $\Var(J_T(\theta_0))$ is evaluated by considering only
the first non-constant term \eqref{first}.

Equation \eqref{first} can be rewritten as
\begin{align*}
  \sum_{i,j,t,\ell,h=1}^T   A_{j\ell th} = \sum_{i,j,\ell=1}^T    \frac{\partial }{\partial \theta_0} 
  \frac{\gamma^{-1}(i-j)}{\sigma_0^2}\gamma^{-1}(0) \frac{\partial }{\partial \theta_0}  \frac{\gamma^{-1}(i-\ell)}{\sigma_0^2}\gamma(j- \ell),
\end{align*}
where $\gamma(j- \ell)=\Gamma_{j\ell}$ and
$\gamma^{-1}(i-j)=\Gamma^{ij}$.  Let $k=i-j$ and $r=j-\ell$.  Without
lose of generality, we assume that $\gamma(h)=0$ if $|h|>T$.  Then the
previous expression and consequently the first term in \eqref{eq_}
simplifies to
\[
  \sum_{k,r=-T}^{T} \left(T-\max\{|k|, |k+r|, |r|\}\right)
  \frac{\partial }{\partial
    \theta_0}\frac{\gamma^{-1}(k)}{\sigma_0^2}\gamma^{-1}(0)
  \frac{\partial }{\partial \theta_0}
  \frac{\gamma^{-1}(k+r)}{\sigma_0^2}\gamma(r).
\]

The absolute summability of the autocovariance and the duality
properties of autocorrelation and of its inverse for causal invertible autoregressive-moving average processes (see
Cleveland (1972) \cite{Cleveland:1972}, Chatfield (1979) \cite{Chatfield:1979} and Hosking (1980) \cite{Hosking:1980}) guarantee the
following holds:
\begin{eqnarray}
  \lefteqn{\lim_{T \rightarrow \infty} \sum_{r,k=-T}^T  \frac{(T-\max\{|k|, |k+r|, |r|\})}{T}}\nonumber\\
& & \mbox{}\times \frac{\partial}{\partial \theta_0}\frac{\gamma^{-1}(k)}{\sigma_0^2} \gamma^{-1}(0) 
\frac{\partial }{\partial \theta_0}\frac{\gamma^{-1}(k+r)}{\sigma_0^2}\gamma(r)\nonumber\\
 &=& \sum_{r,k=-\infty}^{\infty}   \frac{\partial}{\partial \theta_0}\frac{\gamma^{-1}(k)}{\sigma_0^2}\gamma^{-1}(0) \frac{\partial }{\partial \theta_0}\frac{\gamma^{-1}(k+r)}{\sigma_0^2} \gamma(r)\nonumber\\
\label{limit}
& = & \sum_{r,k=-\infty}^{\infty}   V(r,k,\theta_0),
\end{eqnarray}
where
$$V(r,k, \theta_0)=\left(\frac{\partial }{\partial
    \theta_0}\frac{\gamma^{-1}(0)}{\sigma^2_0} \right)^2+
\frac{\partial }{\partial \theta_0} \frac{\gamma^{-1}(k)}{\sigma_0^2}
\gamma^{-1}(0) \frac{\partial \gamma^{-1}(k+r)}{\partial \theta_0}
\gamma(r).$$

Combining equations~\eqref{limit} and \eqref{second} we obtain
\begin{equation}\label{var}
  \Var(J_T(\theta_0)) \longrightarrow  V,
\end{equation}
where $V= \sum_{r,k=-\infty}^{\infty} V(r,k,\theta_0)$.

Since $J(\theta_0)$ depends on the $B_{i}$'s, which involve the
sample autocovariance, it follows from the asymptotic normality of the
sample autocovariance of ARMA processes that $J_T(\theta_0)$ is also
asymptotically normal with zero mean and variance $V$.  From
\eqref{grad} and \eqref{var} we obtain the asymptotic normality of
$\widehat{\theta}_T$:
\[
  \sqrt{T}(\widehat{\theta}_T-\theta_0)\xrightarrow{\mathcal{D}}
  N_{m-1} \left(0, K(\theta_0)^{-1} V K^T(\theta_0)^{-1}\right).
\]

\newpage
\begin{table}[htb]
  \caption{Simulation~1.  Estimated mean  ($Est.$),  asymptotic standard deviation ($sd$), 
    and asymptotic relative efficiency ($\ARE$) of estimators  of the parameter $\phi$
    in the $\AR(1)$ model, for $n=200$, $T=50$, and varying values of $\phi$.  
    We denote by $\widehat{\phi}$ the maximum likelihood estimate, by $\widehat{\phi}_{\pl}$ 
    the pairwise likelihood estimate, and by $\widehat{\phi}_{\sHT}$ and 
    $\widehat{\phi}_{\sHW}$ the total and the matrix Hyv\"arinen estimates, respectively.}
  \label{tab1}       
  \centering
  \vspace{0.2cm}
  \scalebox{0.67}{
    \renewcommand{\arraystretch}{1.5}
    \setlength{\tabcolsep}{3.1pt}
    \begin{tabular}{lllllllllllllllll}
      \hline\noalign{\smallskip}
      \multicolumn{1}{l}{} & \multicolumn{1}{l}{} & \multicolumn{2}{c}{$\widehat{\phi}$} &  \multicolumn{1}{l}{} & \multicolumn{3}{c}{$\widehat{\phi}_{\pl}$} &  \multicolumn{1}{l}{} & \multicolumn{3}{c}{$\widehat{\phi}_{\sHT}$} &  \multicolumn{1}{l}{} & \multicolumn{3}{c}{$\widehat{\phi}_{\sHW}$}\\
      \cline{3-4} \cline{6-8} \cline{10-12} \cline{14-16} 
      \multicolumn{1}{c}{$\phi$} &   \multicolumn{1}{l}{}  & \multicolumn{1}{c}{$Est.$} & \multicolumn{1}{c}{$sd$} &   \multicolumn{1}{l}{} & \multicolumn{1}{c}{$Est.$} & \multicolumn{1}{c}{$sd$} & \multicolumn{1}{c}{$\ARE$} &  \multicolumn{1}{l}{} &  \multicolumn{1}{c}{$Est.$} & \multicolumn{1}{c}{$sd$} & \multicolumn{1}{c}{$\ARE$} &  \multicolumn{1}{l}{}  &  \multicolumn{1}{c}{$Est.$} & \multicolumn{1}{c}{$sd$} & \multicolumn{1}{c}{$\ARE$}\\
      \noalign{\smallskip}\hline\noalign{\smallskip}
      $-0.9$ & $ $ & $-0.8997$ & $0.0041$ & $ $ & $-0.8997$ & $0.0045$ & $0.8625$ & $ $ & $-0.9008$ & $0.0150$ & $0.0738$ & $ $ & $-0.9004$ & $0.0244$ & $0.0278$\\
      $-0.8$ & $ $ & $-0.8000$ & $0.0059$ & $ $ & $-0.7999$ & $0.0064$ & $0.8340$ & $ $ & $-0.8007$ & $0.0146$ & $0.1602$ & $ $ & $-0.8007$ & $0.0236$ & $0.0613$\\
      $-0.7$ & $ $ & $-0.7002$ & $0.0071$ & $ $ & $-0.7001$ & $0.0079$ & $0.8087$ & $ $ & $-0.7007$ & $0.0139$ & $0.2599 $ & $ $ & $-0.7005$ & $0.0226$ & $0.0979$\\
      $-0.6$ & $ $ &  $-0.6002$ & $0.0080$ & $ $ & $-0.6002$ & $ 0.0089$ &  $0.7986$ & $ $ &  
                                                                                             $-0.6008$ & $0.0130$ & $0.3794$ & $ $ & $-0.6008$ & $0.0216$ & $0.1367$ \\
      $-0.5$ & $ $ & $-0.5001$ & $0.0087$ & $ $ & $-0.4999$ & $0.0097$ & $0.8069$ & $ $ &  
                                                                                          $-0.5009$ & $0.0122$ & $0.5060$ & $ $ & $ -0.5011$ & $0.0202$ & $0.1853$ \\
      $-0.4$ & $ $ & $-0.4002$  & $0.0092$ & $ $ & $-0.4000$ & $0.0101$ & $0.8351$ & $ $ &
                                                                                           $-0.4006$ & $ 0.0115$ & $0.6466$ & $ $  & $-0.4001$ &  $0.0184$ & $0.2505$ \\
      $-0.3$ & $ $ & $-0.2997$ & $0.0096$ & $ $ & $-0.2997$ & $0.0102$ & $0.8808$ & $ $ & $-0.2998$ & $0.0109$ &  $0.7773$ & $ $ & $-0.2995$ & $0.0164$ & $0.3438$\\
      $-0.2$ & $ $ & $-0.2003$ &  $0.0099$ & $ $ &  $-0.2002$ & $0.0102$ & $0.9347$ & $ $ & $-0.2005$ &  $0.0104$ & $0.8991$ & $ $ & $-0.2007$ & $0.0143$ & $0.4780$\\
      $-0.1$ & $ $ & $-0.0997$ & $0.0100$ & $ $ & $-0.0997$ & $0.0101$ & $0.9813$ & $ $ & $-0.0997$ & $0.0102$ & $0.9776$ & $ $ & $-0.0999$ & $0.0125$ & $0.6493$\\
      $0$ & $ $ & $0.0002$ & $0.0101$ & $ $ & $0.0002$ & $0.0101$ & $0.9998$ & $ $ & $0.0002$ & $0.0101$ & $1.0077$ & $ $ & $0.0003$ & $0.0117$ & $0.7401$\\
      $0.1$ & $ $ & $0.1005$ & $0.0100$ & $ $ & $0.1005$ & $0.0101$ & $0.9810$ & $ $ & $0.1005$ & $0.0101$ & $0.9810$ & $ $ & $0.1007$ & $0.0125$ & $0.6506$ \\
      $0.2$ & $ $ & $0.1997$ & $0.0099$ & $ $ & $0.1997$ & $0.0102$ & $0.9350$ $ $ & $ $ &  $0.1998$ 
                                                           & $0.0104$ & $0.8980$ & $ $ & $0.1995$ & $0.0143$ &  $0.4802$  \\
      $0.3$ & $ $ & $0.2997$ & $0.0096$ & $ $ & $0.2997$ & $0.0102$ & $0.8808$ & $ $ & $0.2998$ & $0.0109$ & $0.7774$ & $ $ & $0.2995$ & $0.0164$ & $0.3433$\\                                              
      $0.4$ & $ $ & $0.3993$ & $0.0092$ & $ $ & $0.3993$ & $0.0101$ & $0.8355$ & $ $ & $0.3997$ & $0.0115$ & $0.6451$ & $ $ & $0.3995$ & $0.0184$ & $0.2506$\\
      $0.5$ & $ $ & $0.5002$ & $0.0087$ & $ $ & $0.5003$ & $0.0097$ & $0.8071$ & $ $ & $0.5006$ & $0.0122$ & $0.5077$ & $ $ & $0.5004$ & $0.0201$ & $0.1867$\\
      $0.6$ & $ $ &  $0.5997$ & $0.0080$ & $ $ &  $0.5997$ & $0.0089$ & $0.7985$ & $ $ & $0.5998$ & $0.0130$ & $0.3757$ & $ $ & $0.5990$ & $0.0215$ & $0.1376$\\
      $0.7$ & $ $ & $0.6992$ & $0.0071$ &  $ $ & $0.6992$ & $0.0079$ & $0.8087$ & $ $ & $0.6997$ & $0.0138$ & $0.2630$ & $ $ & $0.6993$ & $0.0227$ & $0.0977$\\
      $0.8$ & $ $ &  $0.8001$ & $0.0058$ & $ $ & $0.8001$ & $0.0064$ & $0.8343$ & $ $ & $0.8006$ & $0.0146$ & $0.1605$ & $ $ & $0.8002$ & $0.0235$ & $0.0618$\\
      $0.9$ & $ $ & $0.8998$ & $0.0041$ & $ $ & $0.8998$ & $0.0044$ & $0.8622$ & $ $ & $0.8999$ & $0.0150$ & $0.0734$ &$ $ & $0.8987$ & $0.0244$ & $ 0.0278$\\
      \noalign{\smallskip}\hline
    \end{tabular}
  }
\end{table}  

\begin{table}[htb]
  \caption{Simulation~2.  Estimated mean  ($Est.$),  asymptotic standard deviation ($sd$), and asymptotic relative efficiency ($\ARE$) of estimators  of the parameter $\alpha$ in the $\MA(1)$ model, for $n=200$, $T=50$, and varying values of $\alpha$.  We denote by $\widehat{\alpha}$ the maximum likelihood estimate, by $\widehat{\alpha}_{\pl}$ the pairwise likelihood estimate, and by $\widehat{\alpha}_{\sHT}$ and $\widehat{\alpha}_{\sHW}$ the total and the matrix Hyv\"arinen estimates, respectively.}
  \newpage
  \vspace{0.2cm}
  \label{tab2}       
  \centering
  \scalebox{0.68}{
    \renewcommand{\arraystretch}{1.5}
    \setlength{\tabcolsep}{3.1pt}
    \begin{tabular}{lllllllllllllllll}
      \noalign{\smallskip}\hline
      \multicolumn{1}{l}{} & \multicolumn{1}{l}{} & \multicolumn{2}{c}{$\widehat{\alpha}$} &  \multicolumn{1}{l}{} & \multicolumn{3}{c}{$\widehat{\alpha}_{\pl}$} &  \multicolumn{1}{l}{} & \multicolumn{3}{c}{$\widehat{\alpha}_{\sHT}$} &  \multicolumn{1}{l}{} & \multicolumn{3}{c}{$\widehat{\alpha}_{\sHW}$}\\
      \cline{3-4} \cline{6-8} \cline{10-12} \cline{14-16} 
      \multicolumn{1}{c}{$\alpha$} &   \multicolumn{1}{c}{}  & \multicolumn{1}{c}{$Est.$} & \multicolumn{1}{c}{$sd$} &   \multicolumn{1}{c}{} & \multicolumn{1}{c}{$Est.$} & \multicolumn{1}{c}{$sd$} & \multicolumn{1}{c}{$\ARE$} &  \multicolumn{1}{c}{} &  \multicolumn{1}{c}{$Est.$} & \multicolumn{1}{c}{$sd$} & \multicolumn{1}{c}{$\ARE$} &  \multicolumn{1}{c}{}  &  \multicolumn{1}{c}{$Est.$} & \multicolumn{1}{c}{$sd$} & \multicolumn{1}{c}{$\ARE$}\\
      \noalign{\smallskip}\hline\noalign{\smallskip}
      $-0.9$ & $ $ & $-0.8998$ & $0.0055$ & $ $ & $-0.8996$ & $0.0167$ & $0.1064$ & $ $ & $-0.8999$ & $0.0064$ & $0.7208$ & $ $ & $-0.8993$ & $0.0074$ & $0.5471$\\
      $-0.8$ & $ $ & $-0.7997$ & $0.0066$ & $ $ & $-0.7996$ & $0.0176$ & $0.1390$ & $ $ & $-0.7998$ & $0.0075$ & $0.7566$ & $ $ & $-0.7992$ & $0.0091$ & $0.5177$\\
      $-0.7$ & $ $ & $-0.6997$ & $0.0075$ & $ $ & $-0.6996$ & $0.0183$ & $0.1692$ & $ $ & $-0.6997$ & $0.0086$ & $0.7583$ & $ $ & $-0.6993$ & $0.0106$ & $0.5020$\\
      $-0.6$ & $ $ & $-0.6004$ & $0.0083$ & $ $ & $-0.6005$ & $0.0182$ & $0.2080$ & $ $ & $-0.6007$ & $0.0095$ & $0.7553$ & $ $ & $-0.6003$ & $0.0119$ & $0.4878$\\
      $-0.5$ & $ $ & $-0.5004$ & $0.0089$ & $ $ & $-0.4999$ & $0.0169$ & $0.2757$ & $ $ & $-0.5007$ & $0.0101$ & $0.7646$ & $ $ & $-0.5002$ & $0.0129$ & $0.4743$\\                                                                                                                                                                                   
      $-0.4$ & $ $ & $-0.4000$ & $0.0093$ & $ $ & $-0.3997$ & $0.0148$ & $0.3984$ & $ $ & $-0.4003$ & $0.0104$ & $0.8038$ & $ $ & $-0.4001$ & $0.0136$ & $0.4713$\\
      $-0.3$ & $ $ & $-0.3003$ & $0.0097$ & $ $ & $-0.3000$ & $0.0126$ & $0.5905$ & $ $ & $-0.3006$ & $0.0105$ & $0.8527$ & $ $ & $-0.3006$ & $0.0139$ & $0.4838$\\
      $-0.2$ & $ $ & $-0.2000$ & $0.0099$ & $ $ & $-0.2002$ & $0.0111$ & $0.7926$ & $ $ & $-0.2001$ & $0.0104$ & $0.9119$ & $ $ & $-0.1999$ & $0.0135$ & $0.5408$\\
      $-0.1$ & $ $ & $-0.1003$ & $0.0101$ & $ $ & $-0.1004$ & $0.0103$ & $0.9456$ & $ $ & $-0.1004$ & $0.0101$ & $0.9882$ & $ $ & $-0.1006$ & $0.0124$ & $0.6557$\\
      $ 0  $ & $ $ & $ 0.0001$ & $0.0101$ & $ $ & $0.0001 $ & $0.0101$ & $1.0082$ & $ $ & $ 0.0001$ & $0.0101$ & $1.0101$ & $ $ & $ 0.0005$ & $0.0117$ & $0.7429$\\ 
      $ 0.1$ & $ $ & $ 0.1000$ & $0.0101$ & $ $ & $0.1000 $ & $0.0103$ & $0.9526$ & $ $ & $ 0.1001$ & $0.0101$ & $0.9933$ & $ $ & $ 0.0997$ & $0.0124$ & $0.6554$\\
      $ 0.2$ & $ $ & $ 0.2000$ & $0.0099$ & $ $ & $0.2000 $ & $0.0111$ & $0.7932$ & $ $ & $ 0.2000$ & $0.0104$ & $0.9171$ & $ $ & $ 0.1994$ & $0.0135$ & $0.5402$\\
      $ 0.3$ & $ $ & $ 0.2994$ & $0.0097$ & $ $ & $0.2996 $ & $0.0126$ & $0.5853$ & $ $ & $ 0.2994$ & $0.0105$ & $0.8475$ & $ $ & $ 0.2992$ & $0.0139$ & $0.4835$\\
      $ 0.4$ & $ $ & $ 0.4000$ & $0.0093$ & $ $ & $0.4006 $ & $0.0148$ & $0.3979$ & $ $ & $ 0.4000$ & $0.0105$ & $0.7938$ & $ $ & $ 0.3994$ & $0.0137$ & $0.4639$\\
      $ 0.5$ & $ $ & $ 0.5002$ & $0.0089$ & $ $ & $0.5000 $ & $0.0169$ & $0.2760$ & $ $ & $ 0.5004$ & $0.0101$ & $0.7672$ & $ $ & $ 0.5000$ & $0.0129$ & $0.4721$\\
      $ 0.6$ & $ $ & $ 0.6001$ & $0.0083$ & $ $ & $0.6000 $ & $0.0182$ & $0.2075$ & $ $ & $ 0.6001$ & $0.0095$ & $0.7643$ & $ $ & $ 0.5993$ & $0.0119$ & $0.4850$\\
      $ 0.7$ & $ $ & $ 0.6999$ & $0.0075$ & $ $ & $0.6997 $ & $0.0182$ & $0.1707$ & $ $ & $ 0.6999$ & $0.0086$ & $0.7682$ & $ $ & $ 0.6996$ & $0.0106$ & $0.5047$\\
      $ 0.8$ & $ $ & $ 0.7999$ & $0.0066$ & $ $ & $0.7997 $ & $0.0175$ & $0.1402$ & $ $ & $ 0.8000$ & $0.0075$ & $0.7639$ & $ $ & $ 0.7995$ & $0.0091$ & $0.5209$\\
      $ 0.9$ & $ $ & $ 0.8999$ & $0.0055$ & $ $ & $0.8997 $ & $0.0167$ & $0.1072$ & $ $ & $ 0.9000$ & $0.0064$ & $0.7300$ & $ $ & $ 0.8995$ & $0.0074$ & $0.5504$\\
      \noalign{\smallskip}\hline
    \end{tabular}
  }
\end{table}

\begin{table}[htb]
  \caption{Simulation~3.  Estimated mean  ($Est.$),  asymptotic standard deviation ($sd$), and asymptotic relative efficiency ($\ARE$) of estimators  of the parameter $d$ in the $\ARFIMA$ model, for $n=200$, $T=50$, and varying values of $d$.  We denote by $\widehat{d}$ the maximum likelihood estimate, by $\widehat{d}_{\pl}$ the pairwise likelihood estimate, and by $\widehat{d}_{\sHT}$ and $\widehat{d}_{\sHW}$ the total and the matrix Hyv\"arinen estimates, respectively.}
  \vspace{0.2cm}
  \label{tab3}       
  \centering
  \scalebox{0.7}{
    \renewcommand{\arraystretch}{1.5}
    \setlength{\tabcolsep}{3.1pt}
    \begin{tabular}{lllllllllllllllll}
      \noalign{\smallskip}\hline
      \multicolumn{1}{l}{} & \multicolumn{1}{l}{} & \multicolumn{2}{c}{$\widehat{d}$} &  \multicolumn{1}{l}{} & \multicolumn{3}{c}{$\widehat{d}_{\pl}$} &  \multicolumn{1}{l}{} & \multicolumn{3}{c}{$\widehat{d}_{\sHT}$} &  \multicolumn{1}{l}{} & \multicolumn{3}{c}{$\widehat{d}_{\sHW}$}\\
      \cline{3-4} \cline{6-8} \cline{10-12} \cline{14-16} 
      \multicolumn{1}{c}{$d$} &   \multicolumn{1}{c}{}  & \multicolumn{1}{c}{$Est.$} & \multicolumn{1}{c}{$sd$} &   \multicolumn{1}{c}{} & \multicolumn{1}{c}{$Est.$} & \multicolumn{1}{c}{$sd$} & \multicolumn{1}{c}{$\ARE$} &  \multicolumn{1}{c}{} &  \multicolumn{1}{c}{$Est.$} & \multicolumn{1}{c}{$sd$} & \multicolumn{1}{c}{$\ARE$} &  \multicolumn{1}{c}{}  &  \multicolumn{1}{c}{$Est.$} & \multicolumn{1}{c}{$sd$} & \multicolumn{1}{c}{$\ARE$}\\
      \noalign{\smallskip}\hline\noalign{\smallskip}
      $0.01$ & $ $ & $0.0121$ &  $0.0059$ & $ $ & $0.0101$ &   $0.007$ &  $0.7015$ & $ $ &  $0.0101$ & $0.0059$ & $0.9866$ & $ $ & $0.0105$ & $0.0087$ & $0.4537$\\
      $0.05$ & $ $ & $0.0526$ & $0.0062$ & $ $ & $0.0499$ & $0.0067$ & $0.8585$ & $ $ & $0.0499$ & $0.0065$ & $0.9257$ & $ $ & $0.0504$ & $0.0117$ & $0.2827$\\ 
      $0.1$ & $ $ & $0.1034$ & $0.006$ & $ $ & $0.0997$ & $0.0062$ & $0.9593$ & $ $ & $0.1$ & $0.0067$ & $0.8241$ & $ $ & $0.1001$ & $0.0128 $ & $0.223$\\
      $0.15$ & $ $ & $0.1545$ & $ 0.0052$ & $ $ & $0.15$ & $ 0.0054$ & $0.9258$ & $ $ & $0.1503$ & $0.0058$ & $0.8226$ & $ $ & $0.1497$ & $0.0108 $ & $0.2349$\\
      $0.20$ & $ $ & $0.2041$ & $0.0038$ & $ $ & $0.1999$ & $0.0043$ & $0.8061$ & $ $ & $ 0.2$ & $0.0041$ & $0.8809$ & $ $ & $0.1997$ & $0.0077$ & $0.2475$\\
      $0.25$ & $ $ & $0.2587$ & $0.0021$ & $ $ & $0.2499 $ & $0.0026$ & $ 0.6173$ & $ $ & $0.2499$ & $0.0021$ & $0.9339$ & $ $ & $0.2495$ & $0.0038$ & $0.3005$\\
      $0.3$ & $ $ & $0.3032$ & $0$ & $ $ & $0.3$ & $0.0009$ & $0$ & $ $ & $0.3$ & $0.0001$ & $0$ & $ $ & $0.3$ & $0.0043$ & $0$\\ 
      \noalign{\smallskip}\hline
    \end{tabular}
  }
\end{table}

\end{document}